\documentclass[a4paper,11pt]{article}

\usepackage{float}
\usepackage{graphicx}
\usepackage{authblk}
\usepackage{subcaption}  % For subfigure support
\usepackage{amsmath, amsthm, amssymb}
\usepackage{bm}
\usepackage[x11names]{xcolor} % Add this for enhanced color options
\usepackage[backend=biber,style=numeric,sorting=none,citestyle=numeric-comp]{biblatex}
\usepackage{hyperref}
\hypersetup{
    colorlinks=true,
    linkcolor=black,
    citecolor=black,
    urlcolor=black,
    filecolor=black
}

\newcommand{\chhh}{c_{\text{hhh}}}
\newcommand{\ctth}{c_{\text{tth}}}
\newcommand{\ctthh}{c_{\text{tthh}}}
\newcommand{\cggh}{c_{\text{ggh}}}
\newcommand{\cgghh}{c_{\text{gghh}}}

\newcommand{\cH}{c_{\text{H}}}
\newcommand{\cHBox}{c_{\text{H} \Box }}
\newcommand{\ctH}{c_{\text{tH}}}

\newcommand{\ctG}{c_{\text{tG}}}

\newcommand{\mhh}{m_{hh}}
\newcommand{\mh}{m_{h}}

\newcommand{\pT}{p_{\text{T}}}
\newcommand{\pTh}{p_{\text{T}}^{h}}
\newcommand{\pThh}{p_{\text{T}}^{hh}}
\newcommand{\cosTheta}{ \lvert \cos{\theta^*} \rvert}
\newcommand{\etah}{\eta^{h}}

\title{Di-Higgs and Effective Field Theory:\\ Signal Reweighting Beyond $\mhh$}
\date{}

\author[a]{L. Cadamuro \thanks{\href{mailto:luca.cadamuro@cern.ch}
    {luca.cadamuro@cern.ch}}}
\author[b]{T. Ingebretsen Carlson \thanks{\href{mailto:tom.ingebretsen-carlson@fysik.su.se}
    {tom.ingebretsen-carlson@fysik.su.se}}}
\author[b]{J.~Sj\"olin \thanks{\href{mailto:sjolin@fysik.su.se}
    {sjolin@fysik.su.se}}}
\affil[a]{Universit\'e Paris-Saclay, CNRS/IN2P3, IJCLab, France}
\affil[b]{Stockholm University, Department of Physics, Sweden}
\begin{document}
%\linenumbers

\begin{titlepage}
\maketitle

%% Author Information
%\textrm{\large
%    J\"orgen~Sj\"olin$^{1,}$\footnote{\href{mailto:sjolin@fysik.su.se}{sjolin@fysik.su.se}},
%    Luca Cadamuro$^{2,}$\footnote{\href{mailto:luca.cadamuro@cern.ch}{luca.cadamuro@cern.ch}}
%    and Tom Ingebretsen Carlson$^{1,}$\footnote{\href{mailto:tom.ingebretsen-carlson@fysik.su.se}{tom.ingebretsen-carlson@fysik.su.se}}
%}\\[0.3cm]

%\textit{%
%    $^1$ Department of Physics, Stockholm University, Stockholm, Sweden \newline
%    $^2$ Universit\'e Paris-Saclay, CNRS/IN2P3, IJCLab, Orsay, France
%}

\begin{abstract}
\noindent
Di-Higgs ($hh$) production is crucial for probing the Higgs boson self-interaction and understanding the electroweak phase transition. Deviations from Standard Model predictions in $hh$ gluon-gluon fusion (ggF) can be systematically parameterized using effective field theories (EFT), such as Higgs Effective Field Theory (HEFT) and Standard Model Effective Field Theory (SMEFT), as long as the conditions of the EFT are fulfilled. This note presents an improved EFT signal reweighting method that addresses the limitations of approaches that rely solely on the invariant mass of the di-Higgs system $\mhh$, which fail to capture variations in kinematic variables such as the Higgs boson transverse momentum. We show that these limitations are particularly pronounced in scenarios involving strong destructive interference. The proposed method is developed for both EFTs for $hh$ ggF at $\sqrt{s} = 13$ and $13.6$ TeV at next-to-leading order in QCD. We demonstrate a reweighting technique that combines more than one EFT reference sample while incorporating multiple key variables of ggF. These enhancements improve accuracy across phase space, particularly in capturing variation of the Higgs boson transverse momentum. The method employs a convex combination of the reference samples, with weights parametrized by a distance measure, to achieve a more precise reweighting. In principle, this approach can be extended to other processes, provided that suitable reference samples and distance measure are carefully chosen.

\end{abstract}
\end{titlepage}

\tableofcontents
\clearpage

    \section{Introduction}

Di-Higgs production at the Large Hadron Collider (LHC) is of significant interest because it provides a direct probe of the parameter $\lambda$ which governs the strength of the Higgs boson self-interaction. Measuring $hh$ production ultimately aims to shed light on the shape of the Higgs potential, and deepen our understanding of  electroweak phase transition.

In the Standard Model (SM), the Higgs boson self-interaction is determined by its mass and the Fermi coupling constant \cite{MICCO2020100045}. As a result, $hh$ production is fully predicted within the SM, both in terms of its total cross section and differential distributions. In this note we focus on the dominant $hh$ production mode at the LHC, gluon-gluon fusion (ggF), where the two Higgs bosons are produced via a loop-induced mechanism mediated by top quarks. The $hh$ production cross section through ggF is approximately three orders of magnitude smaller than that of single Higgs production \cite{Ajjath:2022kpv, Baglio:2018lrj, Baglio:2020wgt, Borowka:2016ehy, Chen:2019fhs, Chen:2019lzz,Dawson:1998py,deFlorian:2013jea,deFlorian:2015moa,Grazzini:2018bsd,Grigo:2015dia}, making it challenging to observe.

Importantly, deviations from the SM predictions in $hh$ production would affect both the total cross section and the differential distributions, providing a potential window into new physics. These deviations, arising from Beyond the SM (BSM) physics, can be parameterized using EFTs, as long as the conditions of the EFTs are fulfilled. In this work, we focus on the two EFTs: the Higgs Effective Field Theory (HEFT) \cite{brivio2016complete, Heinrich:2020ckp} and the Standard Model Effective Field Theory (SMEFT) \cite{grzadkowski2010dimension,Heinrich:2022idm}.

To measure the parameters of an EFT, it is necessary to construct a continuous parameterization of the differential distributions at the LHC as a function of the Wilson coefficients. This can be achieved using a technique called \textit{signal reweighting}, which is the main focus of this note. The method described is particularly useful when the matrix-elements of the process under study are not explicitly available. The reweighting makes use of a continuous function applied to reference Monte-Carlo (MC) samples to retrieve any EFT prediction, offering a computationally efficient alternative to generating multiple MC samples with full detector simulation for different sets of Wilson coefficients. Previously, a reweighting method for ggF and HEFT at next-to-leading-order (NLO) in QCD was developed using only the invariant mass of the Higgs boson pair ($\mhh$) and an SM MC sample as reference sample \cite{Alasfar:2023xpc}. However, using $\mhh$ alone does not, for example, fully capture changes in the transverse momentum of the Higgs bosons.

In this note, we propose an improved $hh$ ggF EFT reweighting method that incorporates additional kinematic variables and reference samples. Specifically, the variables considered are $\mhh$, $|\cos{\theta^*}|$ and the transverse momentum of the $hh$ system ($\pT^{hh}$). Here $\theta^*$ is the polar angle between one of the Higgs bosons and the beam axis in the center-of-mass frame of the $hh$ system. The inclusion of $\mhh$ and $|\cos{\theta^*}|$ is motivated by the fact that these variables fully parameterize ggF $hh$ production at leading-order (LO) \cite{carvalho2016higgs}. Additionally, $\pT^{hh}$ is included to account for effects of jet radiation, which enter the process at NLO. This multivariate approach provides a more accurate reweighting framework, improving the precision of EFT analyses in $hh$ production at the LHC. The signal reweighting is derived for both EFTs at $\sqrt{s}=13$ and $\sqrt{s}=13.6$ TeV. 

In addition, to further improve the reweighting we also use a convex combination of the two reference samples which depend on a distance measure, where both the reference samples and the measure are carefully selected to improve the performance. 
% It should be noted that this technique does not incorporate the reference samples into the parametrization, but rather applies the reweighting to both samples which are combined via a phase-space dependent weighting factor.  

\section{Setup}

The reweighting method is derived for HEFT and SMEFT, for the interactions deemed relevant in ggF. For SMEFT the following dimension-6 operators are considered \cite{Heinrich:2022idm, SMEFTpowhegctGImplementation}:
\begin{equation}
\begin{split}
\Delta\mathcal{L}_{\text{Warsaw}} &= \frac{C_{\text{H}\Box}}{\Lambda^2} (\phi^{\dagger} \phi)\Box (\phi^{\dagger } \phi) + \frac{C_{\text{H}}}{\Lambda^2} (\phi^{\dagger}\phi)^3  \\& 
+ \frac{C_{\text{tH}}}{\Lambda^2} \left(\phi^{\dagger}{\phi}\bar{q}_L\tilde{\phi}\, t_R + \text{h.c.}\right) \\& 
+ \frac{C_{\text{tG}}}{\Lambda^2} (\bar q_L\sigma^{\mu\nu}T^aG_{\mu\nu}^a\tilde{\phi} \,t_R +{\rm h.c.})\,. 
\\& + \frac{C_{\text{HG}}}{\Lambda^2} \phi^{\dagger} \phi G_{\mu\nu}^a G^{\mu\nu,a}.
\label{eq:warsaw}
\end{split}
\end{equation}
SMEFT can be expressed in different bases, and here the Warsaw basis is used. The parameter $\Lambda$ represents the new physics scale, while the parameters $C_i$ denote the Wilson coefficients.

While for HEFT the included interactions are as follows \cite{Heinrich:2020ckp}:

%\begin{linenomath}
\begin{align}
\label{eq:ewchl}
\Delta{\cal L}_{\text{HEFT}} &=
-m_t\left(\ctth\frac{h}{v}+\ctthh\frac{h^2}{v^2}\right)\,\bar{t}\,t -
\chhh \frac{m_h^2}{2v} h^3\\ &+\frac{\alpha_s}{8\pi} \left( \cggh \frac{h}{v}+
\cgghh \frac{h^2}{v^2}  \right)\, G^a_{\mu \nu} G^{a,\mu \nu}\;. \nonumber
\end{align}
%\end{linenomath}
Here, the $c_i$ are also Wilson coefficients.

To achieve signal reweighting for the EFTs we use the following:
\begin{equation}\label{Eq:RewGeneral}
    w(\mathbf{c}) = \frac{d\sigma(\mathbf{c})/d \Omega }{d\sigma_{\text{ref}}/d \Omega} w_{\text{ref}},
\end{equation}
where $w$ is the MC event weight, $\mathbf{c}$ represents the set of Wilson coefficients and
$d\sigma/d \Omega$ is the differential distribution over some part of phase space, denoted by $\Omega$. Note that previously $\Omega$ was limited to $\mhh$, while in this note we extend $\Omega$ to be $\mhh$, $|\cos{\theta^*}|$ and $\pT^{hh}$. In addition, for $d\sigma_{\text{ref}}$ and $w_{\text{ref}}$ `ref', corresponds to the reference sample, which is used in order to retrieve the predicted EFT event weight $w(\mathbf{c})$. One possible approach is to use an SM MC sample as a reference sample \cite{Alasfar:2023xpc}. In this work, we go beyond this approach and use BSM reference samples to enhance the reweighting method.    

We parametrize Equation \ref{Eq:RewGeneral} for ggF by using the polynomial dependence of the Wilson coefficients in the Matrix Elements. For SMEFT, incorporating the interactions defined in Equation \ref{eq:warsaw}, we use the following:

\begin{equation}\label{Eq:PolySMEFT_NLO}
\begin{split}
  \sigma_{\text{S}}(\mathbf{c},\textbf{A}) &=   
    \text{Poly}_{\text{S}}( \mathbf{c}, \textbf{A}) = \\ 
  & A_1 + A_2  c_{\text{H} \Box} + A_3  c_{\text{H} \Box}^2  + A_4  c_{\text{H}} + A_5  c_{\text{H}}^2  + \\ 
  & A_6  c_{\text{tH}} + A_7  c_{\text{tH}}^2  + A_8  c_{\text{HG}} + A_9  c_{\text{HG}}^2  + \\ 
  & A_{10}  c_{\text{H} \Box}  c_{\text{H}} + A_{11}  c_{\text{H}}  c_{\text{tH}} + A_{12}  c_{\text{H}}  c_{\text{HG}} + \\ 
  & A_{13}  c_{\text{H} \Box}  c_{\text{tH}} + A_{14}  c_{\text{tH}}  c_{\text{HG}} + A_{15}  c_{\text{H} \Box}  c_{\text{HG}} + \\ 
  & A_{16}\cHBox c_{\text{tG}} + A_{17}  c_{\text{H}}  c_{\text{tG}} + A_{18}  c_{\text{tH}}  c_{\text{tG}} + \\ 
  & A_{19}  c_{\text{HG}}  c_{\text{tG}} + A_{20}  c_{\text{tG}}.
\end{split}
\end{equation}
Here, $\Lambda$ is set to 1 TeV and the Wilson coefficients are normalized as $c_i = C_i/\Lambda^2$. Only the linear contribution from the coefficient $\ctG$ is included, due to the implementation in the generator used, as described later in this section.

For HEFT the following polynomial is used \cite{Alasfar:2023xpc} to incorporate the interactions in Equation \ref{eq:ewchl}:
\begin{equation}\label{Eq:PolyHEFT_NLO}
\begin{split}
  \sigma_{\text{H}}(\mathbf{c}, \textbf{A}) &=   
    \text{Poly}_{\text{H}}( \mathbf{c}, \textbf{\text{A}}) =  \\  & A_{1} c^{4}_{\text{tth}} + A_{2}c^{2}_{\text{tthh}} +(A_{3}c^{2}_{\text{tth}} + A_{4}c^{2}_{\text{ggh}})c^{2}_{\text{hhh}} + \\ & A_{5}c^{2}_{\text{gghh}}  +  (A_{6}c_{\text{tthh}}+A_{7}c_{\text{tth}}c_{\text{hhh}})c^{2}_{\text{tth}} +\\ & (A_{8}c_{\text{tth}}c_{\text{hhh}} +A_{9}c_{\text{ggh}}c_{\text{hhh}})c_{\text{tthh}} + \\ &  A_{10}c_{\text{tthh}}c_{\text{gghh}} +(A_{11}c_{\text{ggh}}c_{\text{hhh}}  + A_{12}c_{\text{gghh}})c^{2}_{\text{tth}} + \\ & (A_{13}c_{\text{hhh}}c_{\text{ggh}}+A_{14}c_{\text{gghh}})c_{\text{tth}}c_{\text{hhh}} +\\ &
     A_{15}c_{\text{ggh}}c_{\text{gghh}}c_{\text{hhh}} + A_{16}c^{3}_{\text{tth}}c_{\text{ggh}} + \\ & A_{17}c_{\text{tth}}c_{\text{tthh}}c_{\text{ggh}} +A_{18}c_{\text{tth}}c^{2}_{\text{ggh}}c_{\text{hhh}} + \\ & A_{19}c_{\text{tth}}c_{\text{ggh}}c_{\text{gghh}}
 +A_{20}c^{2}_{\text{tth}}c^{2}_{\text{ggh}} + \\&  A_{21}c_{\text{tthh}}c^{2}_{\text{ggh}}  + A_{22}c^3_{\text{ggh}}c_{\text{hhh}} +  A_{23}c^{2}_{\text{ggh}}c_{\text{gghh}}.
\end{split}
\end{equation} 
The subscripts in `S' and `H' in $\sigma_i$ and $\text{Poly}_i$ indicate SMEFT and HEFT. The coefficients of the polynomials, $\mathbf{A}$, are derived by performing a fit to MC simulations at parton level. 

The differential parametrization is obtained by binning the event variables. This results in using the same functional form for the polynomials, but with different \textbf{A}-coefficients depending on the bin as follows:

\begin{equation}
    \frac{d\sigma_i(\mathbf{c})}{d \Omega^j} = \text{Poly}_i(\mathbf{c}, \textbf{A}^j).
\end{equation}
Here, the index $j$ corresponds to the bin in the event variables. As a result, a separate polynomial is required for each bin.

Finally, equation \ref{Eq:RewGeneral} is parameterized as:
\begin{equation}
    w(\mathbf{c}) = \frac{\text{Poly}_i(\mathbf{c}, \textbf{A}^j)}{\text{Poly}_i(\mathbf{c}_{\text{ref}}, \textbf{A}^j)} w_{\text{ref}}.
    \label{Eq:RewWPoly}
\end{equation}
Where $\mathbf{c}_{\text{ref}}$ refers to setting the numerical values of Wilson coefficients in the polynomial equal to the values of the coefficients of the reference MC sample.
 
The MC simulations used in the fits to determine the $\mathbf{A}$-coefficients are generated with codes implemented in the generator {\tt POWHEG-BOX-V2} \cite{powAlioli_2010,StefanoFrixione_2007}. For SMEFT, the code {\tt gghh\_SMEFT} \cite{Heinrich:2022idm, SMEFTpowhegctGImplementation} is used, while for HEFT {\tt gghh} \cite{Heinrich:2017kxx,Heinrich:2019bkc,Heinrich:2020ckp} is utilized. The proton-proton  collisions in the MC are produced at $\sqrt{s}=13$ and $\sqrt{s}=13.6$ TeV and the PDF set used is PDF4LHC15nlo30 \cite{Butterworth:2015oua}. Consequently, two sets of \textbf{A}-coefficients are derived for each EFT, corresponding to the different collision energies. The prediction includes finite top quark mass and NLO QCD corrections, except for terms related to $\ctG$ in SMEFT, where only LO contributions are included, due to its implementation in the generator. The binning used for reweighting and for determining the A-coefficients is shown in Table \ref{tab:bin_edges}. The bin edges for $\mhh$ are chosen to reflect the expected experimental resolution. For $\cosTheta$, only four uniformly spaced bins are used, as $\cosTheta$ is generally quite flat as a function of the Wilson coefficients. The binning of $\pThh$ is finer at low $\pThh$, where most events are expected.

\begin{table}[h]
    \centering
    \caption{Bin edges used for the reweighting and the fit to determine the \textbf{A}-coefficients.}
    \begin{tabular}{c|l }
    Variable & Bin edges \\ \hline
    $\mhh$ [GeV] & 250, 270, 290, 310, 330, 350, 370, 390, 410, 430, 450, 490, 530, 570 \\
    & 610, 650, 700, 750, 800, 850, 900, 1000, 1200, 1400 \\ 
    $\cosTheta$ & 0, 0.25, 0.5, 0.75, 1.0 \\
    $\pThh$ [GeV] & 0, 20, 40, 70, 100, 140, 200, 290, 2500 \\ \hline
    \end{tabular}
    \label{tab:bin_edges}
\end{table}
For HEFT, 63 MC samples with $1.6\times10^5$ events each are used in the fit, while for SMEFT, 70 samples with $1.0\times10^5$ events each are utilized. The number of samples and events is determined by incrementally adding MC samples and performing the fit until the reweighting results remain stable across various differential distributions. The numerical values of the Wilson coefficients for the MC samples are selected randomly within the range $[-4\pi, +4\pi ]$. 

Furthermore, note that {\tt POWHEG-BOX-V2} includes a Sudakov form factor to account for extra radiations. This introduces an exponential dependence on the matrix elements, causing the Wilson coefficient dependence in the process to deviate from a strictly polynomial form for kinematic variables strongly affected by extra jet radiation. However, we assume that the dominant contributions are captured by the polynomials in Equation \ref{Eq:PolySMEFT_NLO} and \ref{Eq:PolyHEFT_NLO}.

\section{Methods for improving the reweighting}

In this section, we outline a series of methods aimed at improving the reweighting of the $hh$ ggF EFT signal compared to the $\mhh$-based reweighting where an SM sample is used as the reference sample. While the $\mhh$-based reweighting is effective in many cases, this approach has notable limitations, particularly in fully capturing variables such as the transverse momentum of the Higgs bosons.

To address these shortcomings, we introduce and evaluate improvements to the reweighting procedure. Each subsection builds upon the preceding one, demonstrating incremental enhancements and their impact on the accuracy of EFT predictions. We begin by assessing the performance of reweighting using only $\mhh$ and progressively incorporate additional techniques, including:
\begin{itemize}
    \item Using alternative reference samples. 
    \item Expanding the phase-space variables in the reweighting procedure.
    \item Using a convex combination based on a distance measure to combine two reference samples.
\end{itemize}

In this section, we focus on the distribution of the average transverse momentum of the Higgs bosons, $\pTh$, and aim to demonstrate improvements in capturing its variations. For brevity, we primarily present distributions for $\cH$ and $\cHBox$ in SMEFT at $\sqrt{s} = 13.6$ TeV, as the reweighting methods exhibit similar behavior and performance for other Wilson coefficients in both SMEFT and HEFT, and performing equally well at $\sqrt{s} = 13.6$ TeV and $\sqrt{s} = 13$ TeV. The numerical values of the Wilson coefficients are chosen to highlight when the $\mhh$-based reweighting can capture the $\pTh$ well and worst case-scenarios where its accuracy is reduced. Furthermore, all distributions are truth-level parton kinematics.  

\subsection{SM reference sample and $\mhh$ (1D SM) }

\begin{figure}[H]
\begin{center}$
\begin{array}{cc}
\includegraphics[width=0.5 \textwidth]{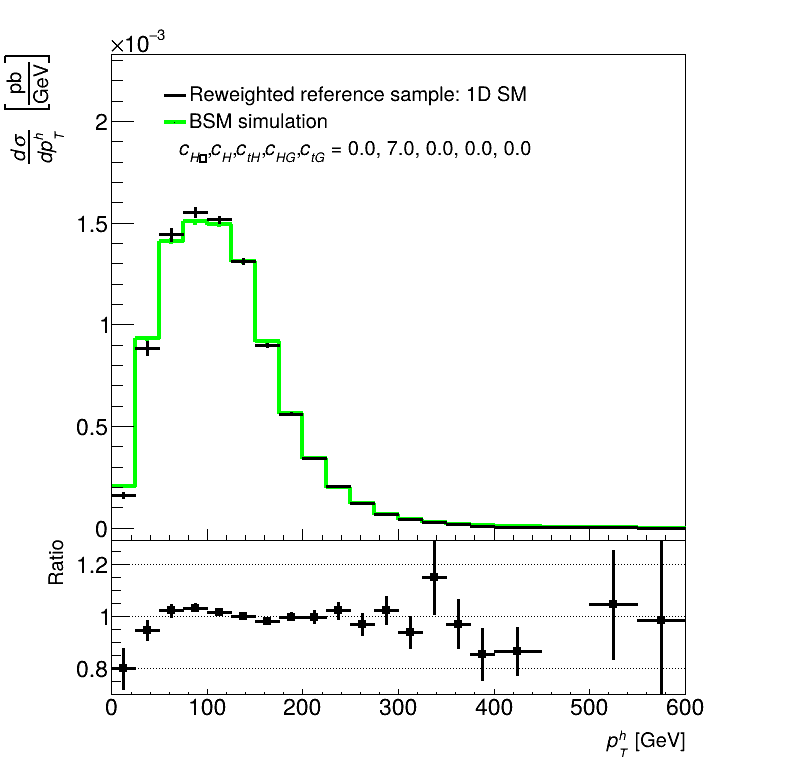} &
\includegraphics[width=0.5 \textwidth]{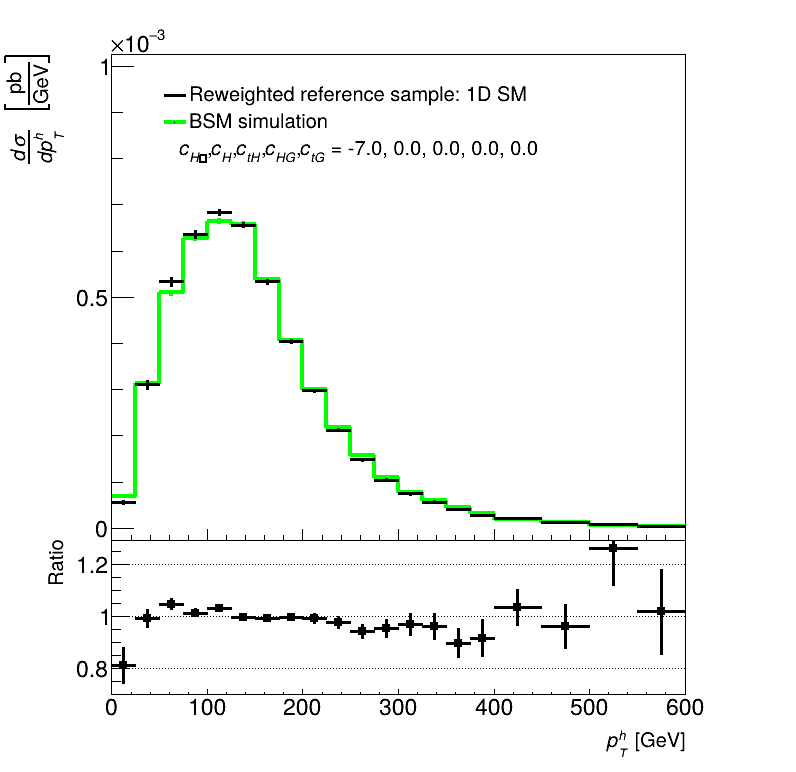} \\
\text{(a)} & \text{(b)} \\[6pt]
\end{array}$
\end{center}
\caption{The truth parton-level $\pTh$ distributions at $\sqrt{s} = 13.6$ TeV predicted by the SMEFT 1D SM reweighting (black dots) are compared to MC simulations (green line). The Wilson coefficient values are chosen such that there is constructive interference between the SM and EFT contributions.}
\label{fig:SMEFTrew_SM_mhh_CI}
\end{figure}

\begin{figure}[H]
\begin{center}$
\begin{array}{cc}
\includegraphics[width=0.5 \textwidth]{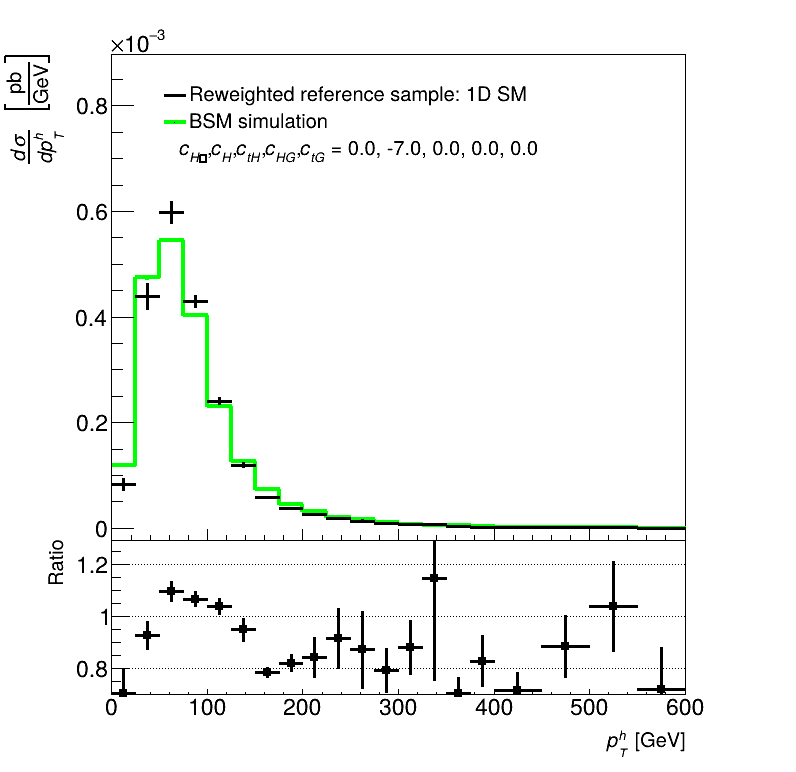} &
\includegraphics[width=0.5 \textwidth]{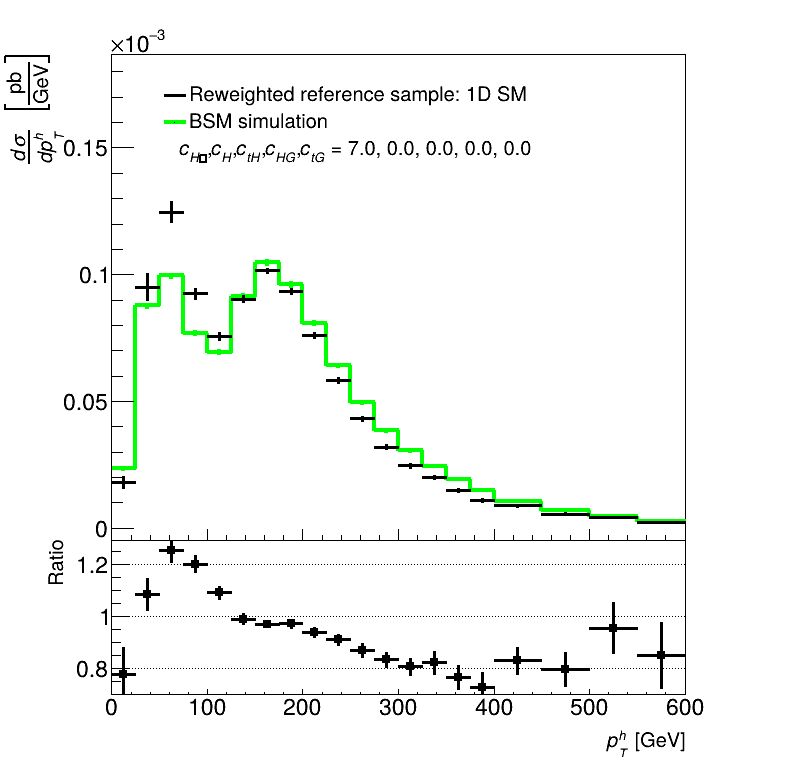} \\
\text{(a)} & \text{(b)} \\[6pt]
\end{array}$
\end{center}
\caption{The truth parton-level $\pTh$ distributions at $\sqrt{s} = 13.6$ TeV predicted by the SMEFT 1D SM reweighting (black dots) are compared to MC simulations (green line). The Wilson coefficient values are chosen such there is destructive interference between the SM and EFT contributions.}
\label{fig:SMEFTrew_SM_mhh_DI}
\end{figure}

In this subsection, the reweighting is performed using an SM reference sample, with $\mhh$ as the phase-space variable. This method will be referred to as `1D SM'. Figure \ref{fig:SMEFTrew_SM_mhh_CI} shows the $\pTh$ distribution when there is constructive interference, for $\cH$ (a) and $\cHBox$ (b). The $\pTh$ distribution is well captured by the reweighting in this scenario. However, Figure \ref{fig:SMEFTrew_SM_mhh_DI} presents the $\pTh$ distribution with destructive interference for $\cH$ (a) and $\cHBox$ (b). In this case, the $\pTh$ distribution is not accurately captured, with discrepancies exceeding 20\% in the tail for $\cH$ and up to 20\% for $\cHBox$. The observation that reweighting methods effectively capture the dependence of $\pTh$ for constructive interference but not for destructive interference is consistent across the Wilson coefficients in SMEFT and HEFT when using only $\mhh$ as the phase-space variable, except for $\cggh$ and $\ctth$ for which these effects are not apparent.

As the destructive interferences seem to be the main source for the discrepancies, we will focus on those values of the Wilson coefficients in the subsequent subsections.

\subsection{EFT reference sample and $\mhh$ (1D EFT)}

\begin{figure}[H]
\begin{center}$
\begin{array}{cc}
\includegraphics[width=0.5 \textwidth]{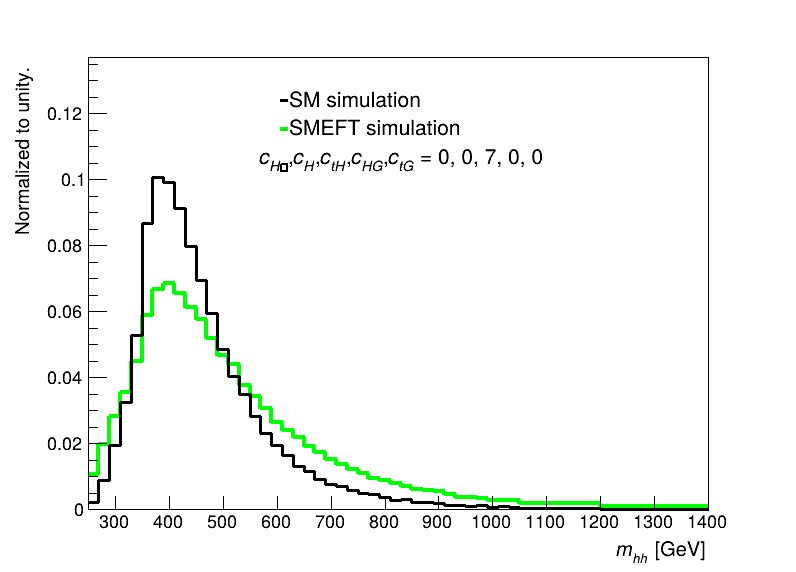} &
\end{array}$
\end{center}
\caption{The truth parton-level $\mhh$ distribution for $\ctH = 7$ (green line) and the SM (black line) normalized to unity with $\sqrt{s} = 13.6$ TeV.}
\label{fig:mhh_cth7}
\end{figure}

Here, we use $\ctH = 7$ as the reference sample and $\mhh$ as the phase-space variable for reweighting, and this method is referred to as `1D EFT'. This reference sample is chosen because it has a broad $\mhh$ distribution, which is expected to provide better coverage of the $hh$ ggF EFT phase space compared to the SM. A comparison of the $\mhh$ distribution normalized to unity for the SM (black line) and $\ctH = 7$ (green line) is shown in Figure \ref{fig:mhh_cth7}. As illustrated, $\ctH = 7$ exhibits a broader $\mhh$ distribution than the SM.

\begin{figure}[H]
\begin{center}$
\begin{array}{cc}
\includegraphics[width=0.5 \textwidth]{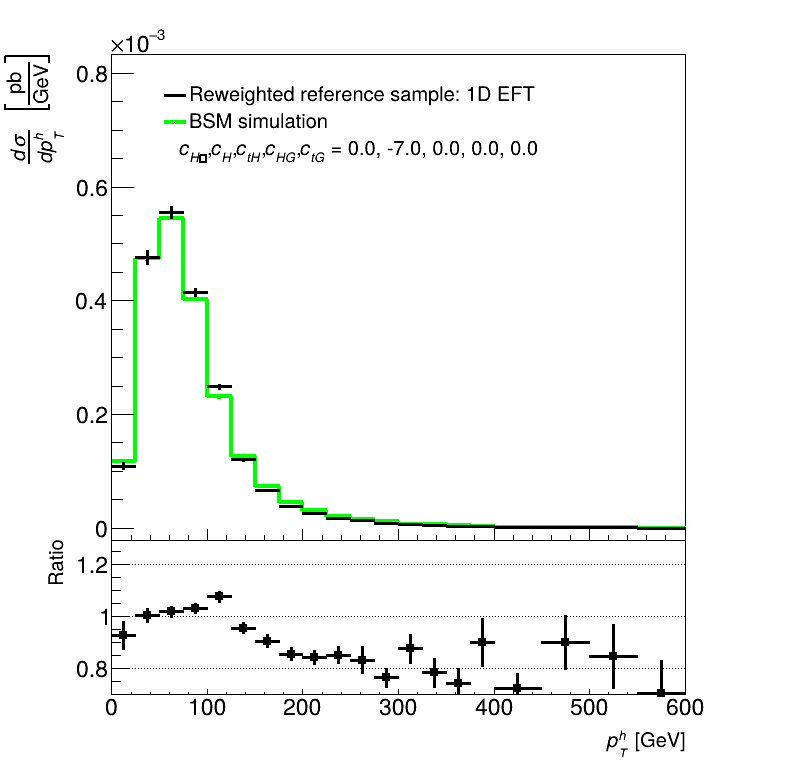} &
\includegraphics[width=0.5 \textwidth]{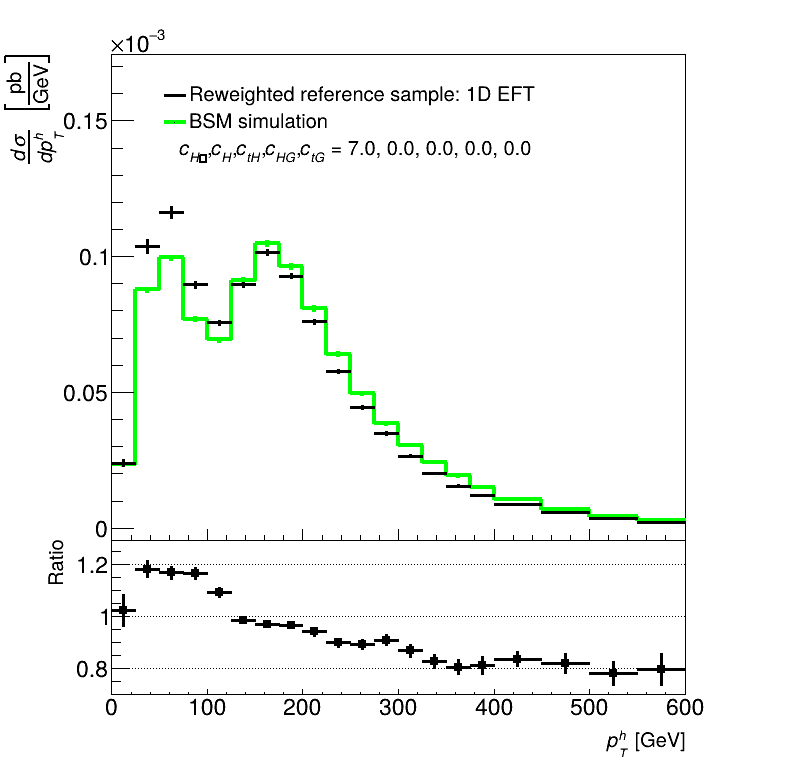} \\
\text{(a)} & \text{(b)} \\[6pt]
\end{array}$
\end{center}
\caption{The truth parton-level $\pTh$ distributions at $\sqrt{s} = 13.6$ TeV predicted by the SMEFT 1D EFT reweighting (black dots) are compared to MC simulations (green line). The Wilson coefficient values are chosen such that there is destructive interference between the SM and EFT contributions.}
\label{fig:SMEFTrew_cth7_mhh_DI}
\end{figure}

In Figure \ref{fig:SMEFTrew_cth7_mhh_DI} the performance of the reweighting is shown for the case of destructive interference. It is shown that the reweighting with the EFT reference sample has less statistical fluctuations compared to using the SM sample (shown in Figure \ref{fig:SMEFTrew_SM_mhh_DI}). However, it is apparent that the method need further improvements to capture the variation in $\pTh$.

\subsection{EFT reference sample and $\mhh$, $\pThh$ and $\cosTheta$ (3D EFT)}\label{subsec:rewmhhcospthh}

\begin{figure}[H]
\begin{center}$
\begin{array}{cc}
\includegraphics[width=0.5 \textwidth]{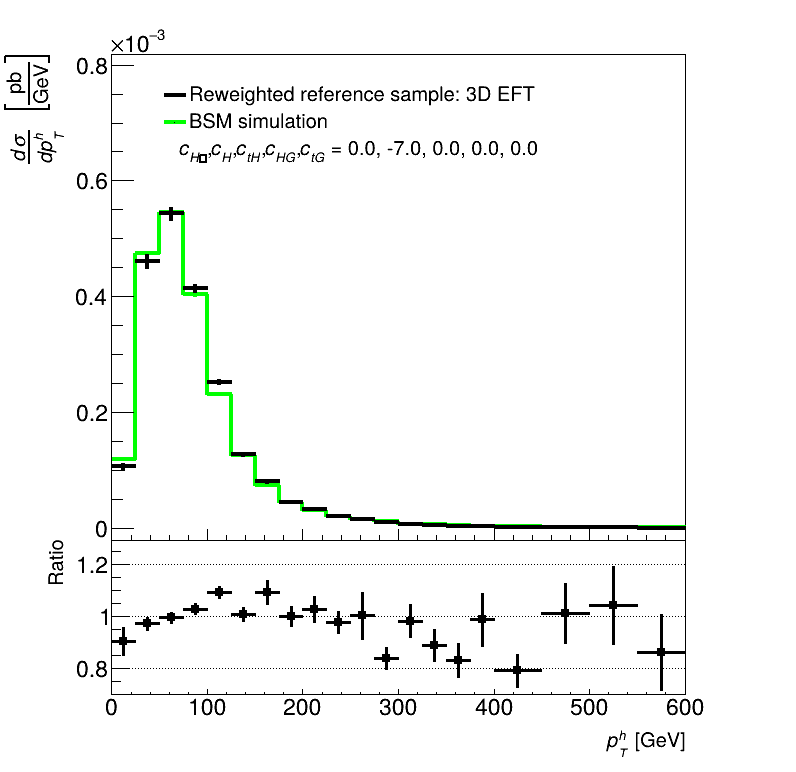} &
\includegraphics[width=0.5 \textwidth]{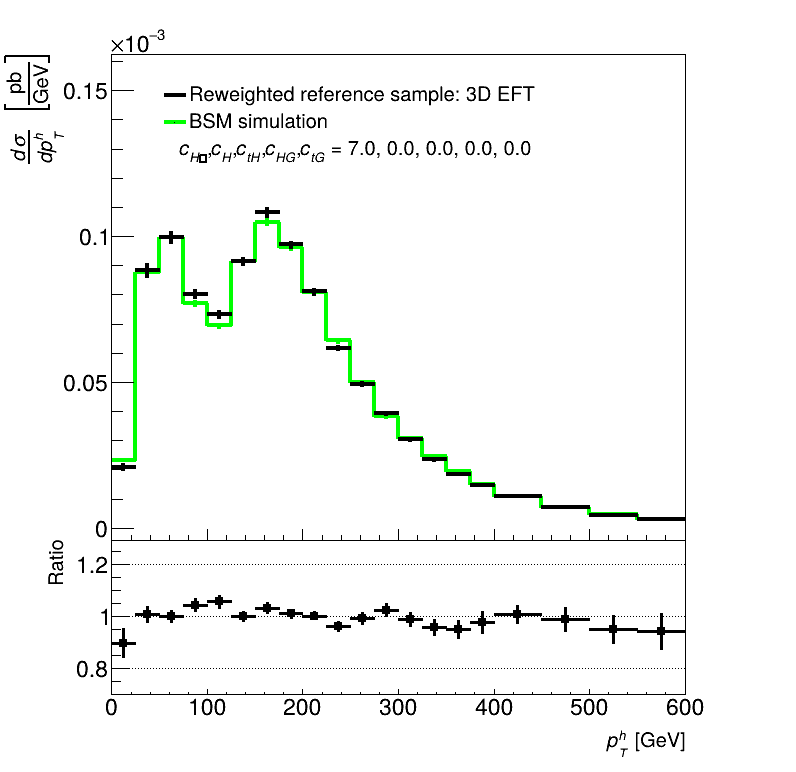} \\
\text{(a)} & \text{(b)} \\[6pt]
\end{array}$
\end{center}
\caption{The truth parton-level $\pTh$ distributions for $\sqrt{s} = 13.6$ TeV, as predicted by the SMEFT 3D EFT reweighting (black dots), are compared to MC simulations (green line).}
\label{fig:SMEFTrew_cth7_Allvars_DI}
\end{figure}

In this subsection, we use $\ctH = 7$ as the reference sample and $\mhh$, $\pThh$, and $\cosTheta$ as the phase-space variables. The method will be referred to as `3D EFT'. Figure \ref{fig:SMEFTrew_cth7_Allvars_DI} shows that, for $\cHBox$ (b), the method effectively captures the $\pTh$ distribution. When compared to Figure \ref{fig:SMEFTrew_cth7_mhh_DI}, one can conclude that this method improves the description of variations in $\pTh$. This behavior generally holds for all Wilson coefficients in SMEFT and HEFT across various values of the coefficients.

\begin{figure}[H]
\begin{center}$
\begin{array}{cc}
\includegraphics[width=0.5 \textwidth]{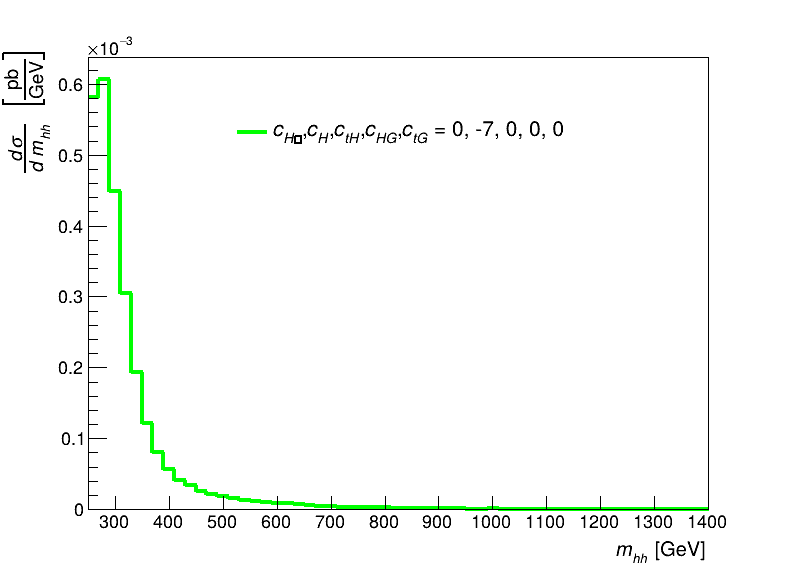} &
\end{array}$
\end{center}
\caption{The truth parton-level $\mhh$ distribution for $\cH = -7$ with $\sqrt{s} = 13.6$ TeV.}
\label{fig:mhh_cpm8p5}
\end{figure}

\begin{figure}[H]
\begin{center}$
\begin{array}{cc}
\includegraphics[width=0.5 \textwidth]{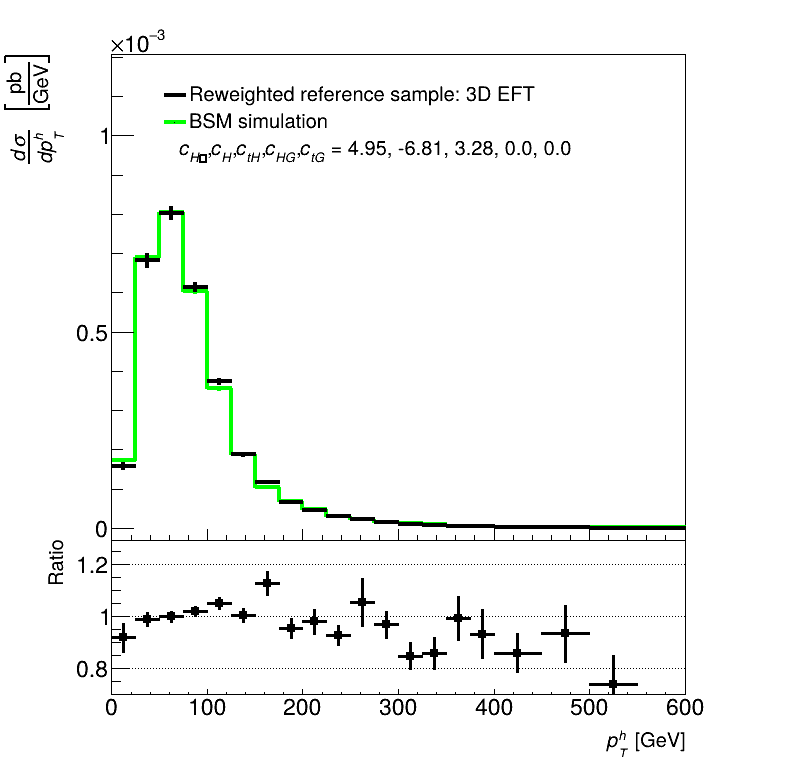} &
\end{array}$
\end{center}
\caption{The truth parton-level $\pTh$ distributions for $\sqrt{s} = 13.6$ TeV, as predicted by the SMEFT 3D EFT reweighting (black dots), are compared to MC simulations (green line).}
\label{fig:SMEFTrew_cth7_Allvars_BM1}
\end{figure}

Certain values of the Wilson coefficient result in an $\mhh$ distribution concentrated near the process threshold, which corresponds to twice the Higgs boson mass ($2m_h$), as can be seen for $\cH = -7$ in Figure \ref{fig:mhh_cpm8p5}. In these scenarios, for example $\cH = -7$ shown in Figure \ref{fig:SMEFTrew_cth7_Allvars_DI} (a), improvements in capturing the $\pTh$ variations are observed compared to Figure \ref{fig:SMEFTrew_cth7_mhh_DI}. However, the tail of the $\pTh$ distribution is not always accurately captured. An example of when the $\pTh$ distribution is not accurately captured can seen in Figure \ref{fig:SMEFTrew_cth7_Allvars_BM1}.

\subsection{Convex combination of reference samples (3D 2xEFT)}\label{subsec:weightedRefSamples}

Here, we present a method to address the discrepancies in the tail of $\pTh$ observed in Section \ref{subsec:rewmhhcospthh} and Figure \ref{fig:SMEFTrew_cth7_Allvars_BM1}, which generally arise when using the 3D EFT reweighting method for points where the $\mhh$ distribution is concentrated near $2 m_h$. To mitigate these discrepancies, we propose to use a convex combination of the reference samples, which depend on a distance measure. The method will be referred to as `3D 2xEFT'. The idea is that when reweighting to target points which exhibits a $\mhh$ close to $2\mh$ as in Figure \ref{fig:mhh_cpm8p5}, one uses a reference sample which exhibits a similar $\mhh$ distribution. It should also be mentioned that using multiple reference samples can contribute to a better performance of the reweighting by carefully selecting reference samples and distance measure which account for effects not included in the parametrization of the reweighting. 

To achieve the reweighting with the convex combination, the reweighting described in Equation \ref{Eq:RewWPoly} is now adapted to the following:

\begin{equation}
    w(\mathbf{c}) = \alpha(D(\mathbf{c})) \frac{\text{Poly}_i(\mathbf{c}, \mathbf{A}^j)}{\text{Poly}_i(\mathbf{c}_{\text{ref}_1}, \mathbf{A}^j)} w_{\text{ref}_1} 
    + (1 - \alpha(D(\mathbf{c}))) \frac{\text{Poly}_i(\mathbf{c}, \mathbf{A}^j)}{\text{Poly}_i(\mathbf{c}_{\text{ref}_2}, \mathbf{A}^j)} w_{\text{ref}_2}.
    \label{Eq:AffineRewWPoly}
\end{equation}
Where \(\mathbf{c}_{\text{ref}_i}\) and \(w_{\text{ref}_i}\) refer to the numerical values of the Wilson coefficients and the MC weights of the reference samples, respectively. Here, $\cH = -8.5$  is used for reference sample 1 and $\ctH = 7$ reference sample 2. The weighting factor \(\alpha\) is a function of a distance measure \(D(\mathbf{c})\), with the condition that \(\alpha \in [0, 1]\).

The distance measure \(D\) we propose is defined as:
\begin{equation}
    D(\mathbf{c}) = 1 - \frac{1}{\sigma(\mathbf{c})} \int_{2 m_h}^{350} \frac{d \sigma(\mathbf{c})}{ d \mhh} d\mhh.
\end{equation}
This distance measure calculates the expected fraction of events with \(\mhh \geq 350\) GeV, which aims to quantify if the $\mhh$ distribution has most of its events close to the threshold $2 m_h$ (as shown in Figure \ref{fig:mhh_cpm8p5}) or not. In practice, this is evaluated using the ratio \(\text{Poly}_i(\mathbf{c}, \mathbf{A}^j) / \text{Poly}_i(\mathbf{c}, \mathbf{A})\), where \(j\) refers to a bin within \(2m_h \leq \mhh \leq 350\) GeV, and \(\text{Poly}_i(\mathbf{c}, \mathbf{A})\) is a polynomial that predicts the inclusive cross-section.

The weighting factor $\alpha$ is calculated as:
\begin{equation}
    \alpha(D(\mathbf{c})) = \min(1, a \cdot \exp(b\cdot D(\mathbf{c}))).
\end{equation}
Here, \(\min\) is the minimum between the exponential and 1, which ensures that \(\alpha \in [0, 1]\), fulfilling the required condition. The constant \(a\) is arbitrarily chosen, and \(b\) is determined such that \(a \cdot \exp(b \cdot D(\mathbf{c}_{\text{ref}_1})) = 1\). This ensures that reference sample 1 is exclusively used when \(\mathbf{c} = \mathbf{c}_{\text{ref}_1}\). In this note, the numerical values \(a = 20\) and \(b = -13.81\) are used.

\begin{figure}[H]
\begin{center}$
\begin{array}{cc}
\includegraphics[width=0.5 \textwidth]{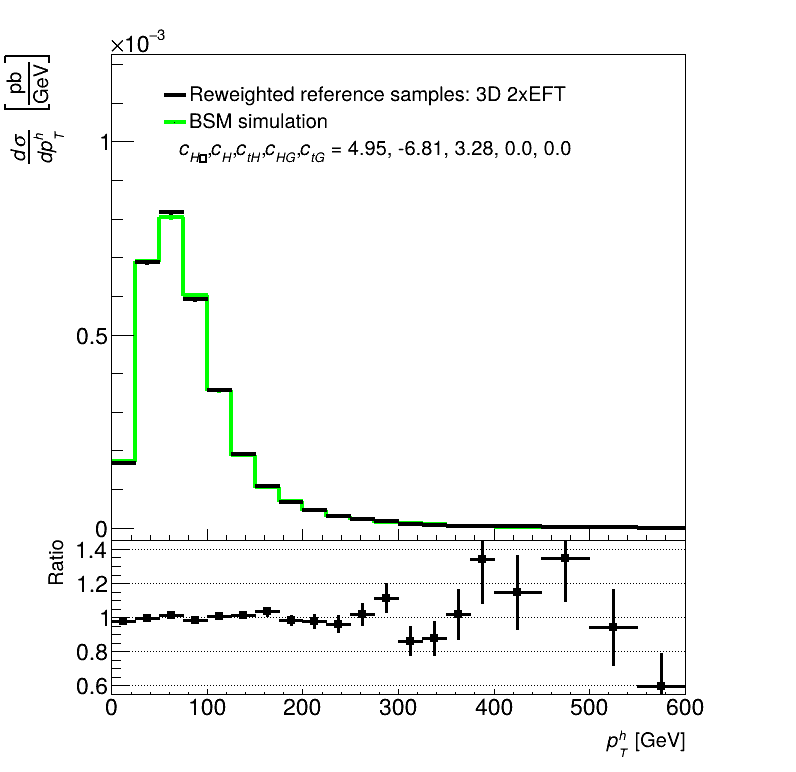}
\end{array}$
\end{center}
\caption{The truth parton-level $\pTh$ distributions for $\sqrt{s} = 13.6$ TeV, as predicted by the SMEFT 3D 2xEFT reweighting (black dots) are compared to MC simulations (green line).}
\label{fig:AffineSMEFTrew_cth7_pth_BM1}
\end{figure}

In Figure \ref{fig:AffineSMEFTrew_cth7_pth_BM1}, the performance of this reweighting method is shown, for the same point as shown in Figure \ref{fig:SMEFTrew_cth7_Allvars_BM1}. Improvements in capturing the dependence of $\pTh$ are observed compared to Figure \ref{fig:SMEFTrew_cth7_Allvars_BM1}. Thus, this method successfully improves the reweighting method in capturing variations in $\pTh$. 

\section{Results and validation}

In this section, the results and validation is presented for the method 3D 2xEFT using convex combination of reference samples, along with the phase-space variables $\mhh$, $\pThh$, and $\cosTheta$ as explained in subsection \ref{subsec:weightedRefSamples}. The reweighting performance is evaluated for various kinematic variables for both HEFT and SMEFT for few selected benchmark (BM) points. Specifically, BM1 is used for SMEFT, while BM7 is considered for HEFT, as defined in Table \ref{tab:SMEFTBMs} for SMEFT \cite{Heinrich:2022idm} and Table \ref{tab:HEFT_BMs} for HEFT \cite{Capozi_2020,Alasfar:2023xpc}. These BM points were chosen because they probe distinct regions of the $\mhh$ spectrum and represent cases where $\mhh$-based reweighting struggled to accurately capture the $\pTh$ dependence.

\begin{table}[h]
    \centering
    \caption{Numerical values of SMEFT $\mhh$ shape benchmark point based on Ref. \cite{Heinrich:2022idm}.}
    \label{tab:SMEFTBMs}
    \begin{tabular}{c | c c c c c}
    \hline BM & $C_{\text{H}\Box}$ & $C_{\text{H}}$ & $C_{\text{tH}}$ & $C_{\text{HG}}$ & $C_{\text{tG}}$ \\ \hline
   1 & 4.95 & -6.81 & 3.28 & 0 & 0\\
\hline
\end{tabular}
\end{table}

\begin{table}[H]
    \centering
    \caption{Numerial values of HEFT $\mhh$ shape benchmark point \cite{Capozi_2020,Alasfar:2023xpc}.}
    \label{tab:HEFT_BMs}
    \begin{tabular}{c|c c c c c}
        \hline
         BM & $c_{\text{hhh}}$ & $c_{\text{tth}}$ & $c_{\text{tthh}}$& $c_{\text{ggh}}$& $c_{\text{gghh}}$\\ \hline
        7 & -0.10 & 0.94 & 1 & 1/6 & -1/6 \\ 
        \hline
    \end{tabular}

\end{table}

The numerical values of the Wilson coefficients we use for the reference samples for SMEFT are indicated in Table \ref{tab:SMEFT_Refs} and HEFT in Table \ref{tab:HEFT_Refs}. The numerical values are chosen by the guiding principle that reference sample 1 should have a $\mhh$-distribution which is centered close to $2 m_h$ (as in Figure \ref{fig:mhh_cpm8p5}) and reference sample 2 should have broad $\mhh$-distributions (as in Figure \ref{fig:mhh_cth7}).

\begin{table}[h]
    \centering
    \caption{Numerical values used for the reference samples for SMEFT.}
    \label{tab:SMEFT_Refs}
    \begin{tabular}{c | c c c c c}
    \hline 
    Ref. sample & $C_{\text{H}\Box}$ & $C_{\text{H}}$ & $C_{\text{tH}}$ & $C_{\text{HG}}$ & $C_{\text{tG}}$ \\ \hline
   1 & 0 & -8.5 & 0 & 0 & 0\\
   2 & 0 & 0 & 7.0 & 0 & 0 \\
\hline
\end{tabular}
\end{table}

\begin{table}[H]
    \centering
    \caption{Numerical values used for the reference samples for HEFT. Where `BM' refers to $\mhh$-shape benchmark points \cite{Capozi_2020,Alasfar:2023xpc}. }
    \label{tab:HEFT_Refs}
    \begin{tabular}{c c |c c c c c}
        \hline
        Ref. sample & BM & $c_{\text{hhh}}$ & $c_{\text{tth}}$ & $c_{\text{tthh}}$& $c_{\text{ggh}}$& $c_{\text{gghh}}$\\ \hline
        1 & 1 & 5.11 & 1.10 & 0 & 0 & 0 \\
        2 & 4 & 2.79 & 0.90 & -1/6 & -1/3 & -1/2 \\
        \hline
    \end{tabular}

\end{table}

\begin{figure}[H]
\begin{center}
\begin{tabular}{cc}
\includegraphics[width=0.5\textwidth]{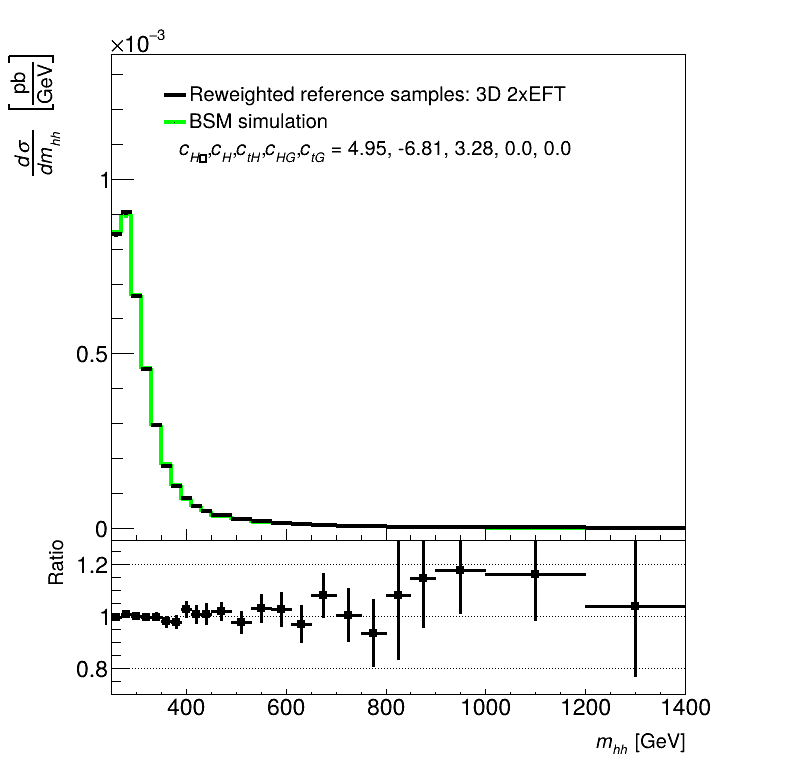} & 
\includegraphics[width=0.5\textwidth]{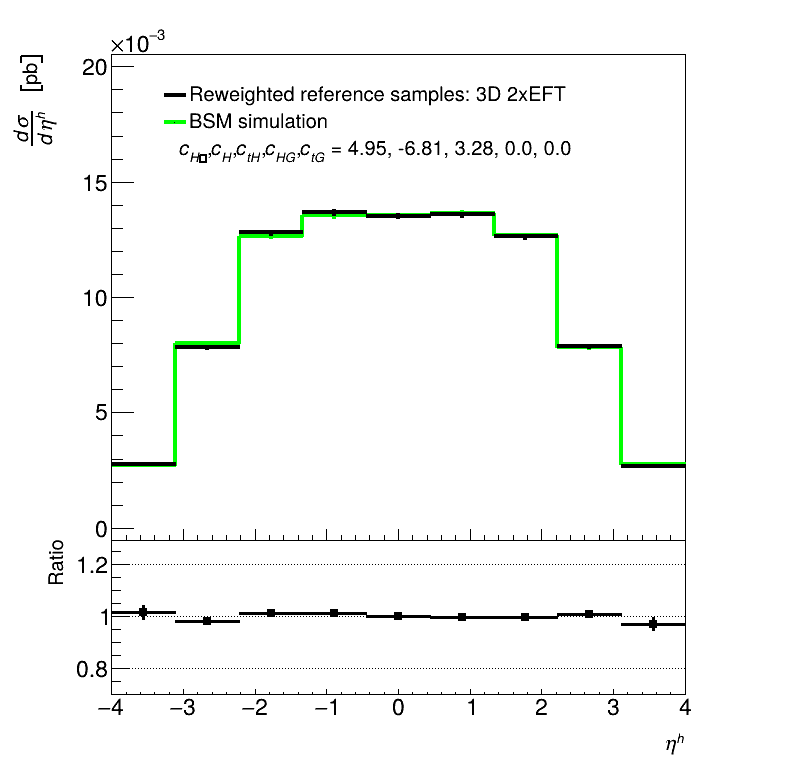} \\
(a) & (b) \\[6pt]
\includegraphics[width=0.5\textwidth]{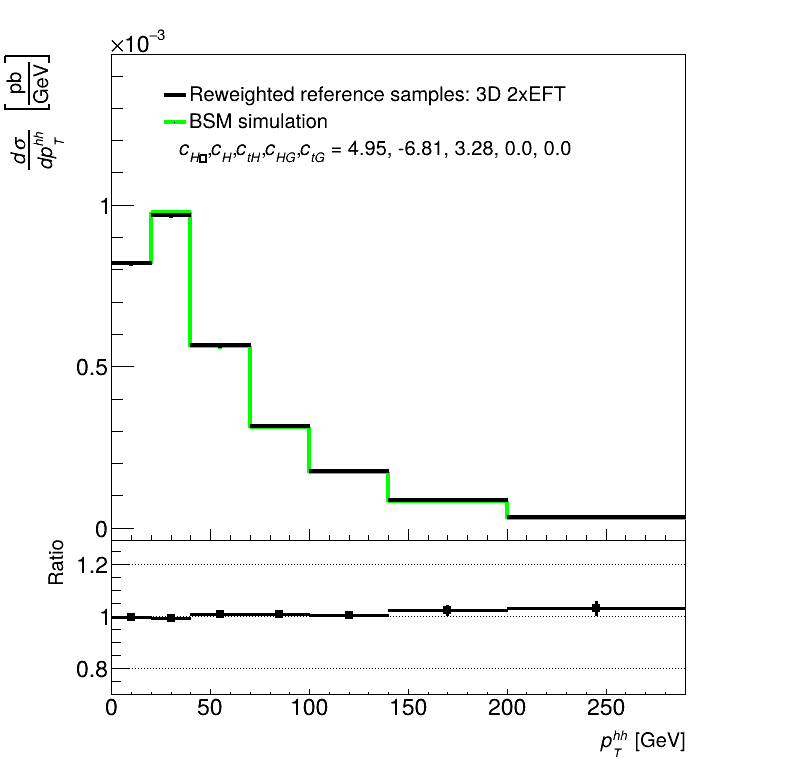} & 
\includegraphics[width=0.5\textwidth]{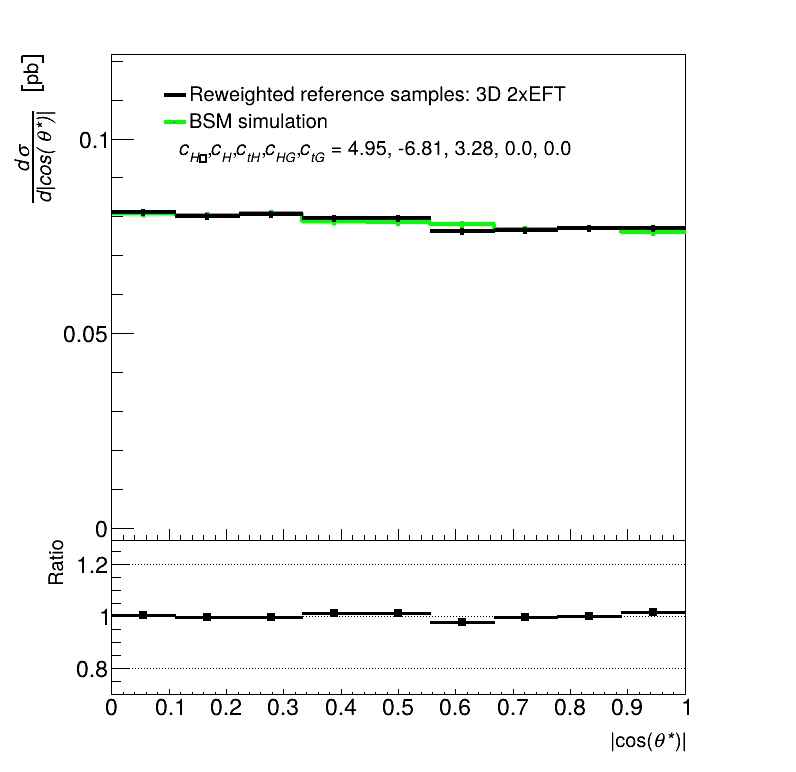} \\
(c) & (d) \\
\end{tabular}
\end{center}
\caption{The truth parton-level distributions for $\mhh$ (a), $\etah$ (b), $\pThh$ (c), and $\cosTheta$ (d) for $\sqrt{s} = 13.6$ TeV, as predicted by the SMEFT 3D 2xEFT reweighting (black dots) compared to MC simulations (green line).}
\label{fig:AffineSMEFTrew_cth7_Allvars_DI_BM1_v1}
\end{figure}

The performance of the SMEFT 3D 2xEFT reweighting is shown for BM1 across the variables $\mhh, \pThh , \left| \cos{\theta^*} \right|$ and the average pseudorapidity of the Higgs bosons, $\etah$ in Figure \ref{fig:AffineSMEFTrew_cth7_Allvars_DI_BM1_v1}. While $\pTh$ for this target point is shown in Figure \ref{fig:AffineSMEFTrew_cth7_pth_BM1}. The plots demonstrate good agreement between the SMEFT 3D 2xEFT reweighting method and MC across all variables.

\begin{figure}[H]
\begin{center}
\begin{tabular}{cc}
\includegraphics[width=0.5\textwidth]{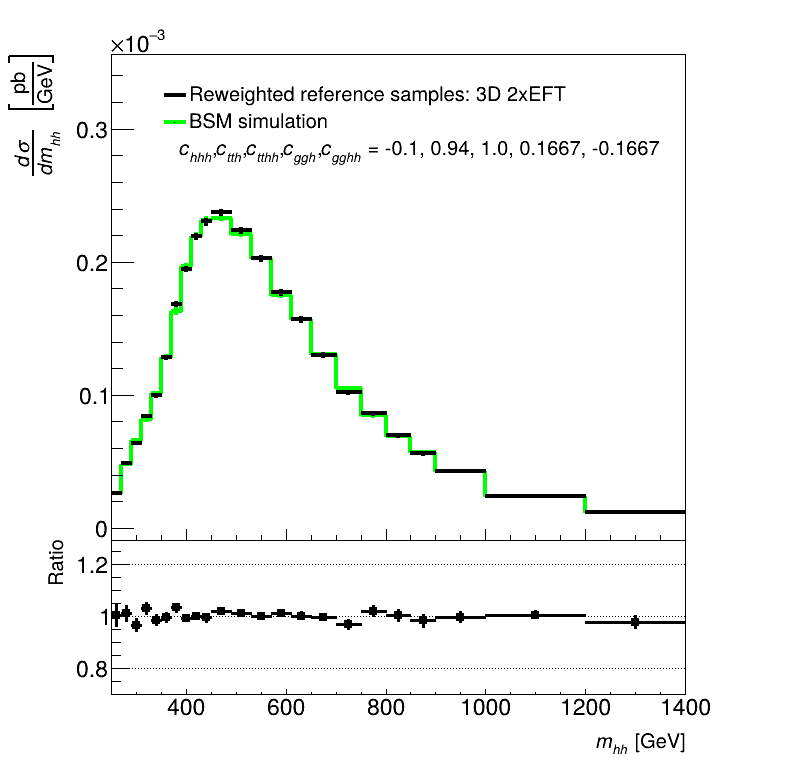} & 
\includegraphics[width=0.5\textwidth]{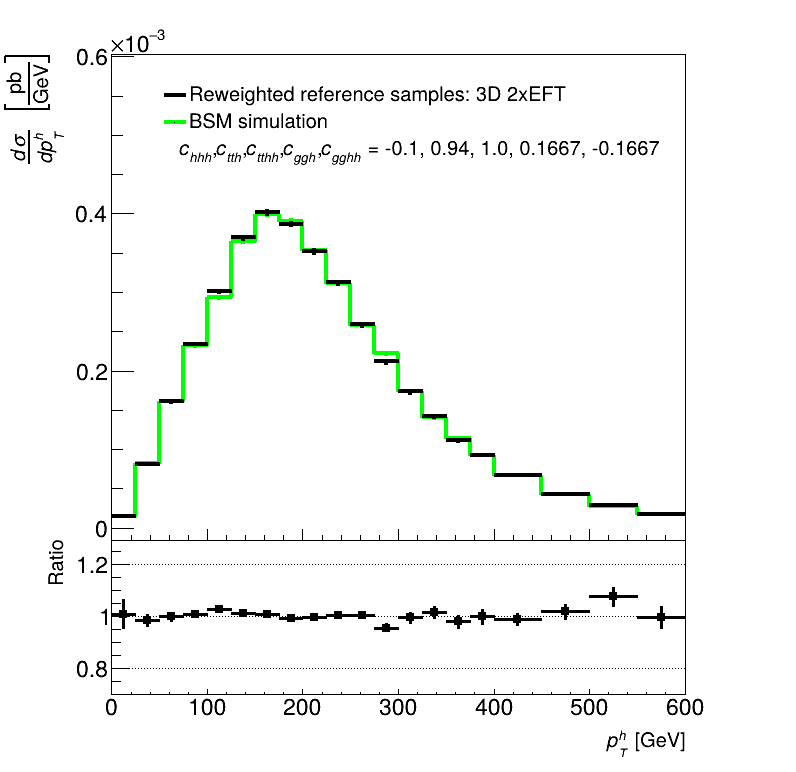} \\
(a) & (b) \\[6pt]
\includegraphics[width=0.5\textwidth]{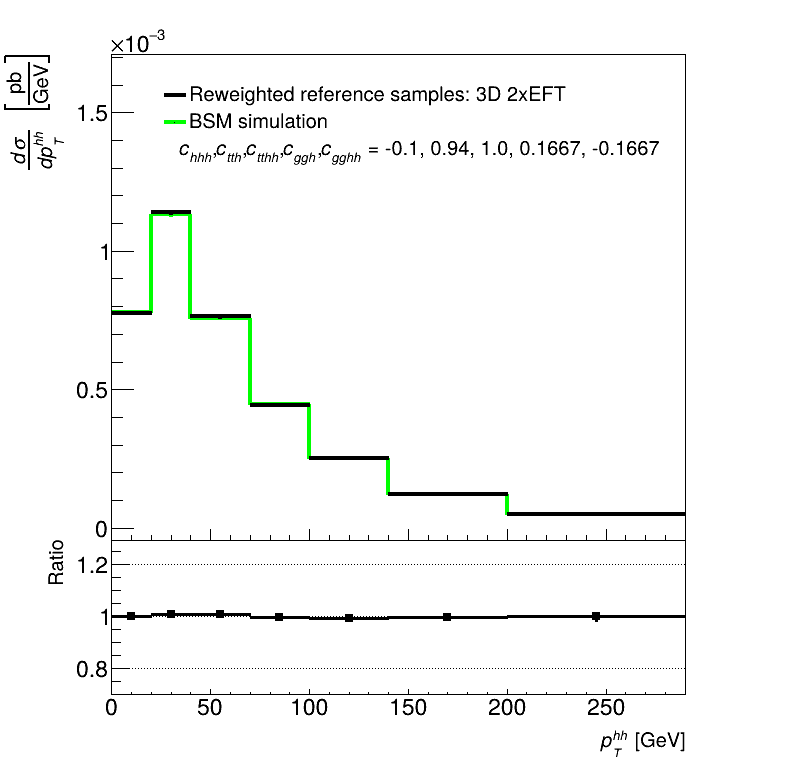} & 
\includegraphics[width=0.5\textwidth]{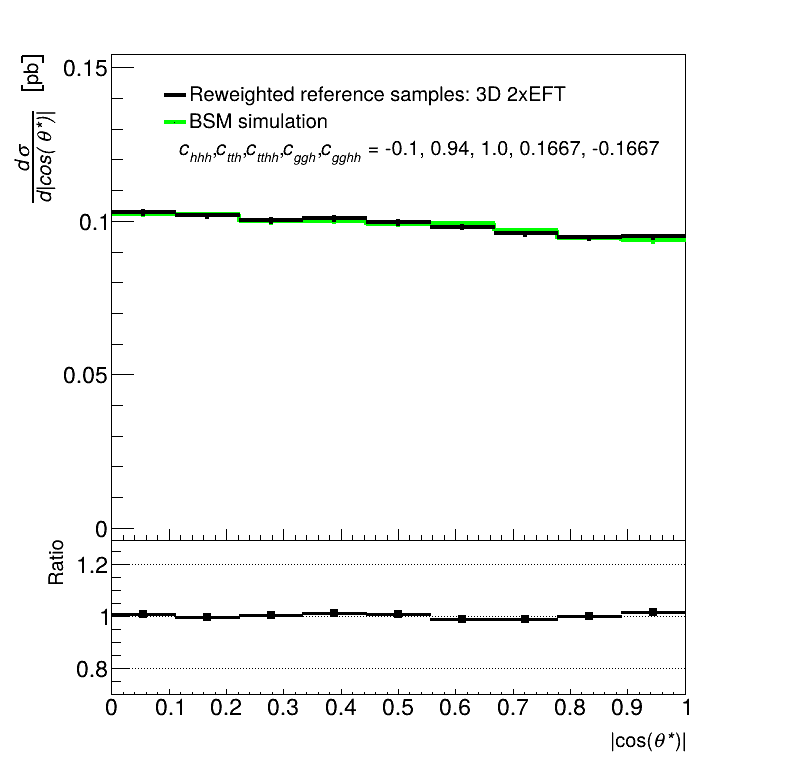} \\
(c) & (d) \\
\end{tabular}
\end{center}
\caption{The truth parton-level distributions for $\mhh$ (a), $\pTh$ (b), $\pThh$ (c), and $\cosTheta$ (d) for $\sqrt{s} = 13.6$ TeV, as predicted by the HEFT 3D 2xEFT reweighting (black dots) compared to MC simulations (green line).}
\label{fig:AffineHEFTrew_Allvars_BM7_v1}
\end{figure}

\begin{figure}[H]
\begin{center}$
\begin{array}{cc}

\includegraphics[width=0.5 \textwidth]{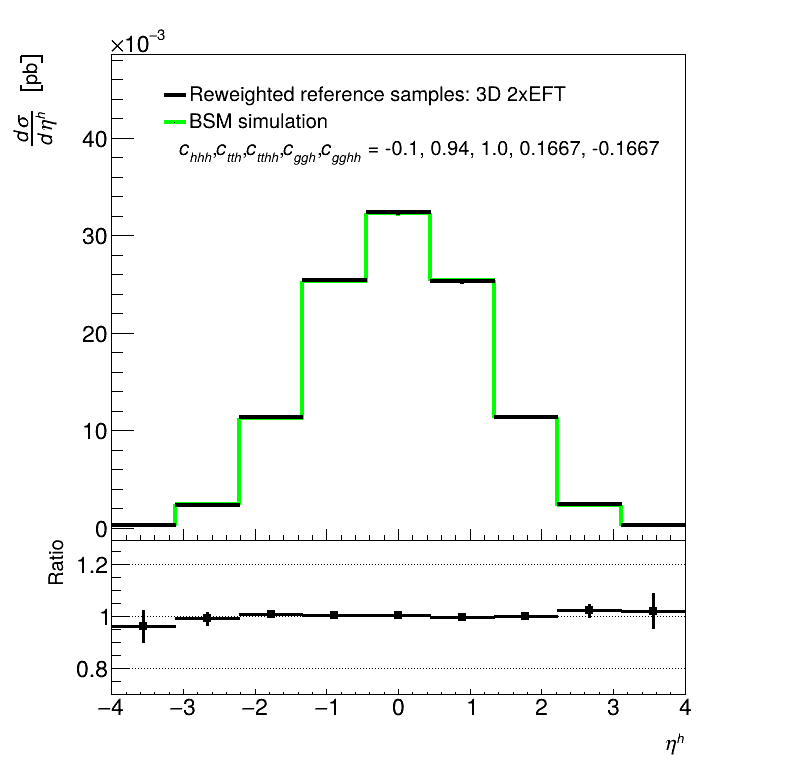} &
\end{array}$
\end{center}
\caption{The truth parton-level $\etah$ distributions for $\sqrt{s} = 13.6$ TeV, as predicted by the HEFT 3D 2xEFT reweighting (black dots) are compared to MC simulations (green line). }
\label{fig:AffineHEFTrew_BM7_Allvars_DI_BM1_v2}
\end{figure}

Similarly, Figure \ref{fig:AffineHEFTrew_Allvars_BM7_v1} and Figure \ref{fig:AffineHEFTrew_BM7_Allvars_DI_BM1_v2} illustrate the performance of the HEFT-based 3D 2xEFT reweighting for BM7 using the same kinematic variables as for SMEFT. Good agreement is observed between the simulation and the reweighting.

\section{Conclusion}

To understand the Higgs potential and the mechanism of the electroweak phase transition, measuring di-Higgs production is crucial. Deviations from the SM predictions in $hh$ ggF production could indicate new physics. EFTs like as HEFT and SMEFT provide a framework to parameterize such deviations, given that the conditions of the EFTs are met, enabling precise comparisons with experimental data and potentially shedding light on physics beyond the SM.

In this note, we present parametrizations of the $hh$ ggF HEFT and SMEFT signals that predict both differential and inclusive cross sections at $\sqrt{s}=13$ and $\sqrt{s}=13.6$ TeV. These parameterizations enables the exploration of the EFT phase space as a continuous function of the Wilson coefficients. Specifically, we employ a method called \emph{signal reweighting}, which typically applies a parameterization to one reference MC sample to predict the EFT behavior for any given set of Wilson coefficient values. While such a method has previously been developed for $hh$ ggF and HEFT using truth-level $\mhh$, we demonstrate that relying solely on $\mhh$ for reweighting fails to fully capture variations in $\pTh$, particularly in scenarios with destructive interference.

This note introduces improvements to the reweighting method beyond $\mhh$. Incremental enhancements are detailed, and the final method with the best performance, referred to as 3D 2xEFT, incorporates reweighting based on truth $\mhh$, $\pThh$, $\cosTheta$ and a convex combination two EFT reference samples based on a distance measure. The variables are chosen as they reflect key aspects of the process: $\mhh$ and $\cosTheta$ completely defines $hh$ ggF production at LO, while $\pThh$ accounts for jet radiation entering the process at NLO. Using an EFT reference sample offers two key advantages: (1) it improves statistical coverage across the EFT phase space, and (2) it allows to capture potential effects that the parametrization might miss, enabling further refinements to the reweighting method. To enhance these benefits, we suggest to use a convex combination of two EFT reference samples, though users may opt for a single reference sample depending on precision requirements. Regardless of the choice, it is essential to assess the closure between the reweighting method and MC for the final discriminating variables in an experimental analysis.

The use of a distance measure to govern the convex combination provides flexibility, allowing users to explore different reference sample configurations and distance measures. This approach can, in principle, be extended beyond $hh$ ggF and applied to other processes, provided that reference samples and distance measures are carefully selected for the specific process and effects under investigation.

\section{Acknowledgements}

This research of T. Ingebretsen Carlson and J. Sjölin is in part supported by the Swedish Research Council grant no.  2023-04654 and the Fysikum HPC Cluster at Stockholm University. The research performed by L. Cadamuro was funded in part by the Agence Nationale de la Recherche (ANR) under the project ANR-22-EDIR-0002.

\printbibliography

\end{document}